\documentclass[12pt]{article}
\usepackage{amsmath,amssymb}
\usepackage{graphicx}
\usepackage{enumerate}
\usepackage{natbib}
\usepackage{booktabs}
\usepackage{bm}
\usepackage{url} 
\usepackage[colorlinks,linkcolor=blue,citecolor=blue,urlcolor=blue]{hyperref}
\usepackage{setspace}
\usepackage{xcolor}
\usepackage{hyperref}
\usepackage{subfigure}
\usepackage{array}
\usepackage{theorem}
\usepackage{algorithm}
\usepackage{algorithmic}
\usepackage{multirow}
\usepackage{comment}
\usepackage{xr}
\usepackage{pbox}
\usepackage{lipsum}
\makeatletter

\newcommand\blfootnote[1]{%
  \begingroup
  \renewcommand\thefootnote{}\footnote{#1}%
  \addtocounter{footnote}{-1}%
  \endgroup
}

\newcommand{\btheta}{{\theta}}
\usepackage{caption}

\def\a{{ a}}

\def\u{{ u}}

\def\x{{ x}}
\def\y{{ y}}
\def\z{{ z}}

\def\I{{ I}}

\def\X{{ X}}
\def\U{{ U}}

\def\W{{ W}}
\def\X{{ X}}
\def\Y{{ Y}}

\def\0{{\bf 0}}
\def\R{{ R}}

\def\1{{ 1}}

\newtheorem{cor}{Corollary}
\newtheorem{prop}{Proposition}
\newtheorem{thm}{Theorem}

\newtheorem{ass}{Assumption}

\begin{document}

\def\spacingset#1{\renewcommand{\baselinestretch}%
{#1}\small\normalsize} \spacingset{1}

\title{\bf A Preferential Latent Space Model for Text Networks}
\date{}
\author{%
{Maoyu Zhang$^1$, Biao Cai$^2$, Dong Li$^3$, Xiaoyue Niu$^4$, Jingfei Zhang$^1$}
\medskip \\
$^\text{1}$ {\normalsize Goizueta Business School, Emory University, Atlanta, GA}\\
$^\text{2}$ {\normalsize Department of Management Sciences, City University of Hong Kong, Hong Kong, China}\\
$^\text{3}$ {\normalsize Department of Statistics and Data Science, 
Tsinghua University, Beijing, China}\\
$^\text{4}$ {\normalsize Department of Statistics, Pennsylvania State University, University Park, PA}\\
}
\maketitle

\blfootnote{Address for correspondence: Jingfei Zhang, Goizueta Business School, Emory University, Atlanta, GA, USA. \url{emma.zhang@emory.edu}; Xiaoyue Niu, Department of Statistics, Pennsylvania State University, University Park, PA, USA. \url{xiaoyue@psu.edu}.}

\bigskip
\begin{abstract}
Network data enriched with textual information, referred to as text networks, arise in a wide range of applications, including email communications, scientific collaborations, and legal contracts. In such settings, both the structure of interactions (i.e., who connects with whom) and their content (i.e., what is communicated) are useful for understanding network relations. Traditional network analyses often focus only on the structure of the network and discard the rich textual information, resulting in an incomplete or inaccurate view of interactions. In this paper, we introduce a new modeling approach that incorporates texts into the analysis of networks using topic-aware text embedding, representing the text network as a generalized multi-layer network where each layer corresponds to a topic extracted from the data. We develop a new and flexible latent space network model that captures how node-topic preferences directly modulate edge formation, and establish identifiability conditions for the proposed model. We tackle model estimation with a projected gradient descent algorithm, and further discuss its theoretical properties. The efficacy of our proposed method is demonstrated through simulations and an analysis of an email network. 
\end{abstract}

\noindent%
\vfill

\newpage
\spacingset{1.9}
\section{Introduction}

Over the past decades, the study of networks has attracted enormous attention, as they provide a natural characterization of complex systems emerging from a wide range of research communities, such as social sciences \citep{borgatti2009network}, business \citep{elliott2014financial} and biomedical research \citep{bota2003gene,zhang2020mixed}. In response to the rising needs in analyzing network data, many statistical network models have been developed, including the exponential random graph model \citep{frank1986markov}, the stochastic block model \citep{wang1987stochastic,nowicki2001estimation}, the random dot product model \citep{athreya2018statistical}, the latent space model \citep{hoff2002latent,ma2020universal,macdonald2022latent}.

The majority of network research to date has focused on networks with binary or weighted edges that characterize the presence or strength of connections between nodes. Meanwhile, networks with textual edges, where an edge between two nodes is a text document, are increasingly common. Examples include email networks, where an edge between two email accounts is an email exchange, and contract networks, where an edge between two firms is a contract. A common approach to analyze such networks is to discard the textual data and use binary or non-negative integer edges to encode the presence or frequency of exchanges. While this simplifies the analysis, it often overlooks important information embedded in the text, such as the topics discussed, the intention of the interaction, or sender/receiver's preferences. This loss of context can lead to an incomplete or inaccurate understanding of the relationships within the network.

A more principled approach to modeling text networks is to incorporate the textual information when modeling the formation of network edges. Towards this goal, texts can be transformed into numerical representations using tools from natural language processing (NLP). Common techniques include Bag-of-Words \citep{joachims1998text}, static embedding methods such as Word2Vec \citep{mikolov2013efficient}, and contextualized embedding methods such as BERT \citep{devlin2018bert} and GPT \citep{achiam2023gpt}. Modern embedding methods such as BERT and GPT capture rich semantic information and usually perform better in predictive tasks. However, dimensions in the embedding space lack clear interpretation, making the resulting representations difficult to interpret within statistical modeling frameworks.

To form interpretable embeddings for each text document, we consider a topic modeling approach that uses transformer-based embeddings, dimensionality reduction, and clustering to generate interpretable and semantically meaningful topics from text data \citep{grootendorst2022bertopic}; see details in Section \ref{sec:real_data}. With the extracted topics, we convert each document $l$ between nodes $i$ and $j$ into a multivariate edge of dimension $K$, denoted as $\y_{ijl} = (y_{ijl}^{(1)}, \ldots, y_{ijl}^{(K)})$, where $y_{ijl}^{(k)}\in\{0,1\}$ indicates the presence of topic $k$ in document $l$. We refer to the resulting network as a \textit{generalized multi-layer network}, where each dimension of the edge corresponds to a distinct topic. This topic-aware embedding greatly enhances model interpretability.

To model this generalized multi-layer network, we propose a new latent space framework that models edge probabilities as a function of latent node positions and node-topic preferences represented via parameter $\W\in\mathbb{R}^{n\times K}$, where $n$ is the number of nodes in the network and $W_{ik}$ characterizes the interest level of node $i$ on topic $k$. The weight $(W_{ik},W_{jk})$ varies across node pairs and topics, allowing the model to flexibly account for varying interest levels from nodes on topics and give direct insights on how they modulate the relationships between nodes. As the number of topics $K$ can be large, we further impose a sparsity assumption on $\W$ to improve model estimability and interpretability. We tackle model estimation with a projected gradient descent algorithm and theoretically derive the error bound of the estimator from each step of the algorithm. A particularly useful output of our model is the direct visualization of node positions in a network, offering insights into the varying roles of nodes across different network layers (see Figures \ref{commonu}-\ref{Wu}).

In summary, our work contributes to both methodology and theory. As to methodology, we develop a modeling framework for a new and understudied class of network data. We consider topic-aware embedding for text associated with each edge in the network, and propose a flexible multi-layer latent space model. The proposed model is able to effectively borrow information across a large number of sparse layers when estimating the latent node positions and also provide direct insights into the heterogeneous node-topic preferences. With respect to theory, we establish an explicit error bound for the projected gradient descent iterations that shows an interesting interplay between computational and statistical errors. Specifically, it demonstrates that as the number of iterations increases, the computational error of the estimates converges geometrically to a neighborhood that is within statistical precision of the unknown true parameter. The theoretical analysis is nontrivial, as it involves alternating gradient descent, orthogonal transformation, identifiability constraints, sparsity, and a non-quadratic loss function. 

Some recent works considered modeling networks with edges that contain textual information. For example, \cite{sachan2012using,bouveyron2018stochastic,corneli2019dynamic,boutin2023embedded} considered Bayesian community-topic models that extended the latent Dirichlet allocation model to incorporate network communities. These works assume nodes in the network form several communities and the focus is to identify the community label of each node. Model estimation in these works is often carried out via Gibbs sampling or variational EM, which may be prohibitive when applied to large networks. In comparison, our goal is to understand the relationships between nodes and we do not impose assumptions on the community structure amongst nodes. 
There is another closely related line of research on modeling {standard multi-layer networks, which are special cases of the generalized multi-layer networks we study, by allowing only one nonzero multivariate edge between two nodes.}
From our data, standard multi-layer networks can be constructed by merging the text documents between a pair of nodes into one. In this case, the edge between nodes $i,j$ on layer $k$ counts the appearance of keyword $k$ in all of the text documents between nodes $i,j$. 
For multi-layer networks, \cite{paul2020spectral,lei2020consistent,jing2021community,agudze2022markov,lei2022bias,lyu2023optimal} and others considered community detection, and 
\cite{gollini2016joint,salter2017latent,d2019latent} studied Bayesian latent space models. 
Recently, \cite{zhang2020flexible,arroyo2021inference} considered multi-layer network models that assume layer-specific scaling, but is unable to capture the varying level of interests from nodes on a specific layer. \cite{wang2023multilayer} introduced multi-layer random dot product graph model and developed a novel nonparametric change point detection algorithm. \cite{macdonald2022latent} proposed a novel latent space model where the latent node positions are concatenations of common position coordinates and layer-specific position coordinates. This model may not work well when there is a large number of sparse layers, as it is challenging to estimate the layer-specific positions in this case. 
We compare with both \cite{zhang2020flexible} and \cite{macdonald2022latent} in simulations and real data analysis. In particular, we find that our proposed method enjoys better prediction accuracy in the analysis of a real email network.

The rest of our paper is organized as follows. Section \ref{sec:model} introduces the preferential latent space model for networks with multivariate edges and Section \ref{sec:est} discusses model estimation. Section \ref{sec:theo} investigates theoretical properties of the estimator from our proposed algorithm. Section \ref{sec:sim} reports the simulation results, and Section \ref{sec:real_data} conducts an analysis of the Enron email corpus data. The paper is concluded with a discussion section.

\section{Preferential Latent Space Model}\label{sec:model}

We start with some notation. Let $[k]=\{1,2\dots,k\}$.
Given a vector $\x\in\mathbb{R}^d$, we use $\Vert\x\Vert_0$, $\Vert\x\Vert_2$ and $\Vert\x\Vert_{\infty}$ to denote the vector $\ell_0$, $\ell_2$ and $\ell_{\infty}$ norms, respectively. Write $\langle{a},{b}\rangle=\sum_i a_ib_i$ for ${a},{b}\in\mathbb{R}^n$. 
For a matrix $\X\in\mathbb{R}^{d_1\times d_2}$, let $\|\X\|_F$ and $\|\X\|_{op}$ denote the Frobenius norm and operator norm of $\X$, respectively, and $\|\X\|_0=\sum_{ij}1(X_{ij}\neq 0)$ denote the number of nonzero entries. We use $\text{Diag}(x_1,\ldots,x_d)$ to denote a $d\times d$ diagonal matrix with diagonal elements $x_1,\ldots,x_d$, and use $\circ$ to denote the Hadamard product. For two positive sequences $a_n$ and $b_n$, write $a_n\precsim b_n$ {or $a_n=O(b_n)$} if there exist $c>0$ and $N>0$ such that $a_n<cb_n$ for all $n>N$, {and $a_n=o(b_n)$ if $a_n/b_n\rightarrow 0$ as $n\rightarrow\infty$}; write $a_n\asymp b_n$ if $a_n\precsim b_n$ and $b_n\precsim a_n$.

Suppose there are $n$ nodes in the network, and between nodes $i,j$, there are $m_{ij}$ document exchanges denoted as $\{\z_{ijl}\}_{l\in[m_{ij}]}$. Each $\z_{ijl}$ is a tokenized document consisting of a list of words. From the corpus $\{\z_{ijl}\}_{i,j\in[n],\,l\in[m_{ij}]}$, we extract a set of $K$ topics (see details in Section \ref{sec:real_data}). 
Correspondingly, each document $\z_{ijl}$ can be represented as a $K$-dimensional vector $\y_{ijl}=(y_{ijl}^{(1)},\ldots,y_{ijl}^{(K)})$, where $y_{ijl}^{(k)}$ characterizes the presence of topic $k$ in document $\z_{ijl}$. In this work, we refer to $y_{ijl}^{(k)}$ as an edge when there is no ambiguity and $\y_{ijl}$ as a {multivariate} edge. 
We focus on undirected binary-valued edges with $y_{ijl}^{(k)}=y_{jil}^{(k)}\in[0,1]$, although our methods and results generalize directly to directed edges and other types of edge values, such as continuous and non-negative integers, using tools in generalized linear models.
We denote the network data we model as ${Y}=\{\Y_{ij}\}_{i,j\in[n]}$, where $\Y_{ij}\in[0,1]^{m_{ij}\times K}$ collects the $m_{ij}$ {multivariate} edges between nodes $(i,j)$, and the $l$th row of $\Y_{ij}$ is the length-$K$ vector, $\y_{ijl}$. If there is no exchange between nodes $i$ and $j$, set $\Y_{ij}=(0,\ldots,0)\in[0,1]^K$. 

\begin{figure}[t!]
    \centering
    \includegraphics[width=0.3\linewidth]{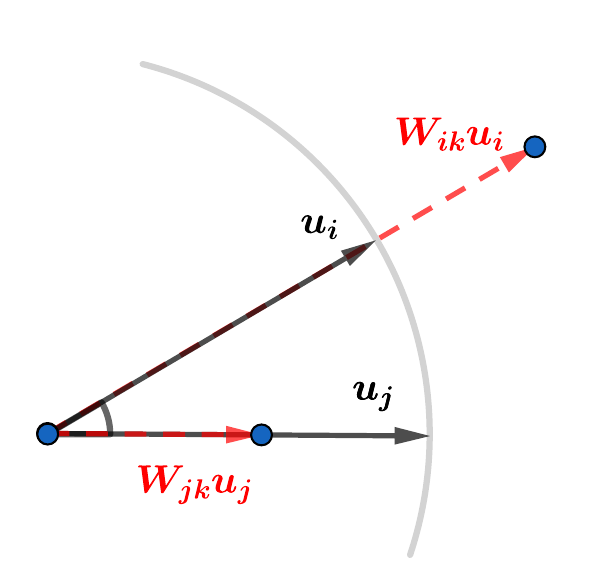}
    \caption{An illustration of the preferential latent space model.}\label{illustration}
\end{figure}

We adopt a conditional independence approach \citep{hoff2002latent} which assumes each node $i$ has a unique latent position $\u_i\in\mathbb{R}^d$. Letting $\U=[ u_1, \ldots, u_n]^{\top}$, the model admits
$$
\text{pr}(Y\mid \U,\btheta)=\prod_{i,j,l,k}\text{pr}(y^{(k)}_{ijl}\mid\u_i,\u_j,\btheta),
$$
where $\btheta$ collects other model parameters to be estimated. Given $\U$ and $\btheta$, we assume that $y^{(k)}_{ijl}$ follows a Bernoulli distribution, with  $\Lambda^{(k)}_{ij}=\text{log odds}(y^{(k)}_{ijl}=1\mid\U,\btheta)$ 
and
\begin{equation}\label{eqn:model}
\Lambda^{(k)}_{ij}=a_i + a_j +(W_{ik} u_i ^{\top})(W_{jk} u_j),
\end{equation}
where $a_i\in\mathbb{R}$ represents the node-specific baseline effect and $W_{ik} \ge 0$ is a weight parameter that quantifies the interest level of node $i$ in topic $k$. 
{In particular, a larger $W_{ik}$ indicates a greater interest of node $i$ in topic $k$.
If either node $i$ or $j$ is uninterested in topic $k$, meaning $W_{ik}=0$ or $W_{jk}=0$, then the log odds of $y^{(k)}_{ijl}=1$ reduces to the baseline level $a_i + a_j$. The parameters $\u_i$ and $\u_j$ are latent node positions, and the angle between them determines the likelihood of edges between nodes $i$ and $j$. When $\u_i$ and $\u_j$ point in the same direction, that is, $\u_i^\top\u_j>0$, the two nodes are more likely to have an edge in any topic $k$. Additionally, if both nodes $i$ and $j$ have strong interests in topic $k$, meaning large $W_{ik}$ and $W_{jk}$, the likelihood of an edge between nodes $i$ and $j$ in topic $k$ is further increased. See Figure \ref{illustration} for an illustration. 
} 

We refer to model \eqref{eqn:model} as the \textit{preferential latent space model} (\texttt{PLSM}) and discuss model identifiability conditions at the end of this section.
In this model, {the probability of an edge between nodes $i,j$ in topic $k$ is determined by $W_{ik}W_{jk}\u_i^\top\u_j$. The multiplied weight $W_{ik}W_{jk}$ varies across node pairs and topics, allowing the model to flexibly account for varying interest levels from nodes on topics. {From this perspective, the proposed model is more flexible than \citet{zhang2020flexible,arroyo2021inference}, where the weight is topic-specific but assumed to be the same across nodes.}
The baseline effect $\a=\left(a_1, \ldots, a_n\right)^{\top}$ and latent positions in $\U$ are shared across all topics, enabling model \eqref{eqn:model} to borrow information across a large number of sparse layers when estimating these parameters.} When $K$ is large, we impose element-wise sparsity on $\W $ to ensure its estimability. This stipulates that each node prefers only a subset of the $K$ topics, or equivalently, each topic is only preferred by only a subset of the nodes. This plausible assumption effectively reduces the number of parameters while enhancing model interpretability. 

The following proposition states a sufficient condition for the identifiability of model \eqref{eqn:model}, and its proof is collected in the supplement.

\begin{prop}[Identifiability]\label{prop1} Suppose two sets of parameters $\left(\a, \W, \U \right)$ and $\left(\a^{\dagger},  {\W}^{\dagger}, \U^{\dagger}\right)$ satisfy the following conditions:\\ 
(1) $\| u_i\|_2=1$, $\|{ u_i^{\dagger}}\|_2=1$ for $i\in[n]$.\\
(2) $\W_{i'k'}=0$, ${\W_{i'k'}^\dagger}=0$ for some $i'\in[n]$, $k'\in[K]$.\\
(3) $\U,\U^{\dagger}\in\mathbb{R}^{n\times d}$ have full column rank.\\
Then, if the following holds for $i,j \in [n], k \in [K]$,
$$
a_i + a_j +(W_{ik} u_i ^{\top})(W_{jk} u_j)=a_{i}^\dagger  + a_{j}^\dagger  +(W_{ik}^\dagger  u_{i}^\dagger{}^{\top})(W_{jk}^\dagger  u_{j}^\dagger ), 
$$
there exists an orthonormal matrix $R \in \mathbb{R}^{d \times d}$ satisfying $\R^{\top} \R=\R \R^{\top}=I_d$, such that
$a^{\dagger}= a$, ${{W}}^{\dagger}= W$ and $\U^{\dagger}=\U \R$.
\end{prop}
Condition (1) in the above proposition is a norm constraint, imposed to ensure $\W$ and $\U$ are identifiable. This condition confines all latent node positions to the unit sphere. Condition (2) assumes at least one entry in $\W$ is zero and this is imposed to ensure $\W$ and $\a$ are identifiable. Condition (3) helps to ensure $\W$ is identifiable.

\section{Estimation}\label{sec:est}
Given network data $Y=\{\Y_{ij}\}_{i,j\in[n]}$, where $\Y_{ij}\in[0,1]^{m_{ij}\times K}$, we aim to estimate model parameters $\a$, $\W$, and $\U$. 
Under model \eqref{eqn:model}, the negative log-likelihood function, up to a constant, can be written as 
\begin{equation}\label{likelihood}
\begin{aligned} & \ell\left( a, {W}, \U \right)  
=  -\sum_{k=1}^K \sum_{i\leq j}^n\sum_{l=1}^{m_{ij}}\frac{1}{m_{ij}}\left\{y^{(k)}_{ijl} \Lambda^{(k)}_{ij}+\log \left(1-\psi(\Lambda^{(k)}_{ij})\right)\right\},\end{aligned}
\end{equation}
where $\psi(x)=1/(1+\exp(-x))$. 
We consider the following optimization problem,
\begin{equation}\label{eq:opt}
\min_{\a \in \mathbb{R}^n, {W} \in \mathbb{R}^{n\times K}_+, \U \in \mathbb{R}^{n\times d}}\ell( a, {W}, \U) \text{, \quad subject to } \|\W\|_0\leq s,
\end{equation}
where $s$ is a tuning parameter that controls the sparsity of $\W$. 
To enforce sparsity and the positivity constraint on $\W$ along the solution path, 
 we employ a truncation operator Truncate($\W, s$) defined as,
$$
[\operatorname{Truncate}(\W, s)]_{ik}= \begin{cases} W_{ik}, & \text { if } (i,k) \in \text{Supp}^+(\W,s),\\ 0, & \text { otherwise, }\end{cases}
$$
for $\W \in \mathbb{R}^{n \times K}$ and $s <nK$. The set $\text{Supp}^+(\W,s)$ denotes the $s$ entries in $\W$ with the largest values. 
To solve \eqref{eq:opt}, we consider a projected gradient descent algorithm that is easy to implement and computationally efficient. Our estimation procedure is summarized in Algorithm \ref{alg1}. 


\begin{algorithm}[!htb]
\begin{spacing}{1.35}
\caption{Projected Gradient Descent Algorithm}
\begin{algorithmic}
\STATE\textbf{Input}: network data $Y$, initial values $\a^{(0)}, W^{(0)},\U^{(0)}$, step sizes $\eta_{\a},\eta_{\W},\eta_{\U}$.
\STATE\textbf{repeat for} $t=0,1,\ldots,$\\
\hspace{0.15in} $\a^{(t+1)}=\a^{(t)}-\eta_{\a} \nabla_{\a} \ell(\a,{\W}^{(t)},\U^{(t)})|_{\a=\a^{(t)}}$;\\
\hspace{0.15in} ${\W}^{(t+1)}={\rm Truncate}\left(\W^{(t)}-\eta_{\W}\nabla_{\W} \ell(\a^{(t)},\W,\U^{(t)})|_{\W=\W^{(t)}},s\right)$;\\
\hspace{0.15in} ${\U}^{(t+1)}=\U^{(t)}-\eta_{\U} \nabla_{\U} \ell(\a^{(t)},{\W}^{(t)},\U)|_{\U=\U^{(t)}}$; normalize rows of ${\U}^{(t+1)}$;\\
\STATE\text{\bf until} the objective function converges.
\STATE\text{\bf Output} $a$, $W$, and $U$
\end{algorithmic}\label{alg1}
\end{spacing}
\end{algorithm}

The parameters $\eta_{\a},\eta_{\W},\eta_{\U}$ control the step sizes in the gradient descent algorithm. Theorem \ref{thm1} provides theoretical conditions on $\eta_{\a},\eta_{\W},\eta_{\U}$ to ensure the algorithm achieves a linear convergence rate. 
In practice, backtracking line search can be implemented for $\eta_{\a},\eta_{\W},\eta_{\U}$ at each step of the iteration to achieve fast convergence.
For the initialization of Algorithm \ref{alg1}, we consider a singular value thresholding based approach \citep{ma2020universal}, which has demonstrated good empirical performance. See Section S1 in the supplement for details. 

The latent dimension $d$ and the sparsity $s$ are two tuning parameters in the proposed model. We select these two parameters using edge cross-validation. 
{Specifically, we divide all indices $\{i,j,l,k\}$'s into $L$ folds and use each fold as a validation set while training the model on the remaining $L-1$ folds. To calculate the cross-validation error on the validation set, we consider the binomial deviance, and the $d$ and $s$ combination with the smallest cross-validation error is selected.

\section{Theoretical Results}\label{sec:theo}
We define the parameter space as
\begin{equation}\label{space}
\begin{aligned}
\Omega_{n,d,K}(M_1)=\bigg\{(\a,\W,\U)\mid&\,\, \|\a\|_{\infty}\leq {M_1}/{4},\,\,  \max_{i}\sum_{k=1}^K(W_{ik})^2\leq {M_1}/{2},\,\,\|\W\|_0<nK,\\
&\| u_i\|_2=1,\,\,\max_{i, j,k}\Lambda_{ij}^{(k)}\leq -(1-C)M_1\bigg\},
\end{aligned}
\end{equation}
where $M_1\ge 0$ is a scalar that may depend on $n$ and $0<C<1$ is a constant.
By the definition of $\Lambda_{ij}^{(k)}$ in \eqref{eqn:model} and combining $\|\a\|_{\infty}\leq {M_1}/{4}$, $\max_{i}\sum_{k=1}^K(W_{ik})^2\leq {M_1}/{2}$ and $\| u_i\|_2=1$ in \eqref{space}, it is straightforward to show that $\max_{i,j,k}|\Lambda_{ij}^{(k)}|\leq M_1$. Hence, for any $(\a,\W,\U) \in \Omega_{n,d,K}(M_1)$, $\Lambda_{ij}^{(k)}$'s are uniformly bounded in $ [-M_1,-(1-C)M_1]$ for any $i,j$ and $k$. 
That is, edge probabilities $\psi({\Lambda_{ij}^{(k)}})$'s are bounded between ${1}/({1+e^{M_1}})$ and ${1}/({1+e^{(1-C)M_1}})$. It is seen that $M_1$ controls the overall sparsity of the network. If, for example, $M_1$ is in the order of $\log(n)-\log\log(n)$, then the average edge probability is in the order of $\log(n)/n$.

Let $(\a^\ast,\W^\ast,\U^\ast)$ be the true parameter, $\sigma_1^\ast\ge\cdots\ge\sigma_d^\ast>0$ be the nonzero singular values of $\U^\ast$ and $s^\ast=\|\W^\ast\|_0$. Write $w_{\max}=\max_k{{w^{(k)}_{\max}}}$, where ${{w^{(k)}_{\max}}}=\max_iW_{ik}^\ast$ is the maximum entry in column $\W^\ast_{.k}$, and $w_{\min}=\min_k{w^{(k)}_{\min}}$, where ${w^{(k)}_{\min}}=\min_{i:W_{ik}^\ast\neq 0}W_{ik}^\ast$ is the minimum nonzero entry in column $\W^\ast_{.k}$. We assume $w_{\max}\asymp {{w^{(k)}_{\max}}}$ and $w_{\min}\asymp {w^{(k)}_{\min}}$ for any $k$. This assumption is made to simplify notations in our analysis, and our results hold under more general conditions on $W_{ik}^\ast$'s but with more involved notations. 
We denote $\bar{m}=(\max_i{1}/{n}\sum_j{1}/{m_{ij}})^{-1}$,
where $\bar{m}$ characterizes the 
average number of edges. To further simplify notation, we assume $\min_{ij} m_{ij}=O(1)$, that is, the minimum number of edges between two nodes is a constant.  

To investigate the computational and statistical properties of iterates from Algorithm~\ref{alg1}, we first introduce an error metric for the iterates from Algorithm~\ref{alg1}. As $\U$ is identifiable up to an orthogonal transformation, for any $\U_1, \U_2\in\mathbb{R}^{n\times d}$, we define a distance measure
\begin{equation*}
\text{dist}(\U_1,\U_2)=\min_{\R:\R\R^\top=\I_d}\|\U_1-\U_2\R\|_F.
\end{equation*} 
Next, we define the error from step $t$ in Algorithm~\ref{alg1} as 
\begin{equation}\label{et2}
e_t=2Kn\|\a^{(t)}-\a^\ast\|_2^2+{\sigma_1^\ast}^2w_{\max}^2\|\W^{(t)}-\W^\ast\|_F^2+K{\sigma_1^\ast}^2w_{\max}^4\text{dist}^2(\U^{(t)}, \U^\ast). 
\end{equation}
We first derive an error bound for $e_t$ in Theorem \ref{thm1}, and then derive error bounds for $\a^{(t)}$, $\W^{(t)}$ and $\U^{(t)}$, respectively, in Corollary \ref{cor1}. We assume the following regularity conditions.

\begin{ass}\label{ass1}
Let $\kappa_{0}=({{\sigma_1^\ast}w_{\max}^2})/({{\sigma_d^\ast}w_{\min}^2})$. Assume initial values $\a^{(0)}$, $\W^{(0)}$ and $\U^{(0)}$ satisfy 
$$
e_0\leq C_{1}K {\sigma_1^\ast}^4w_{\max}^4\kappa_{0}^{-4}e^{-2M_1},
$$ 
for a sufficiently small constant $C_{1}>0$.
\end{ass}
This assumption requires the initial values to be reasonably close to the true parameters. Such assumptions are commonly employed in nonconvex optimizations \citep{lyu2023optimal,zhang2023generalized}. In particular, if $\kappa_0=O(1)$ and $d=O(1)$, then Assumption \ref{ass1} can be simplified to $\|\a^{(0)}-\a^\ast\|_2^2=O(nw_{\max}^4e^{-2M_1})$, $\|\W^{(0)}-\W^\ast\|_F^2=O(Knw_{\max}^2e^{-2M_1})$ and $\text{dist}^2(\U^{(0)},\U^\ast)=O(ne^{-2M_1})$. These assumptions on $\a^{(0)}$, $\W^{(0)}$ and $\U^{(0)}$ are mild. 

\begin{ass}\label{ass2}
Assume the following holds for a sufficiently large constant $C_2>0$,
\begin{equation*}
K{\sigma_d^\ast}^2\geq C_2({w_{\max}^2}/{w_{\min}^4})\max\left\{{n}/{\bar{m}},\log(n)\right\}e^{CM_1}.
\end{equation*}
\end{ass}
This is an assumption on the minimal signal strength $\sigma_d^\ast$, which is the minimum nonzero singular value of $\U^\ast$. It is seen that the signal strength condition weakens as the number of layers $K$ or average number of edges $\bar m$ increases. Also, the signal strength condition becomes stronger as $M_1$ increases, corresponding to sparser networks. 

Next, we are ready to state our main theorem.

\begin{thm}\label{thm1}
Suppose $(\a^{(0)},\W^{(0)},\U^{(0)})$ satisfies Assumption~\ref{ass1}, $(\a^\ast,\W^\ast,\U^\ast)$ is in \eqref{space} and satisfies Assumption~\ref{ass2}, and {$d\kappa_0^8e^{3M_1}=O({\sigma_1^\ast}^2)$}. 
Letting $\eta_\a={\eta}/({4Kn})$, $\eta_{\W}={\eta}/({4{\sigma_1^\ast}^2w_{\max}^2})$, $\eta_{\U}={\eta}/({2K{\sigma_1^\ast}^2w_{\max}^4})$ and $s=\gamma s^\ast$ for $\gamma>1$, the $t$-th step iteration of Algorithm \ref{alg1} satisfies, with probability as least $1-Kn^{-1}$,
\begin{equation*}
e_{t}\precsim \rho^t e_0+\kappa_0^4e^{(1+C)M_1}\left[d\max\left\{{n}/{\bar{m}},\log(n)\right\}+{s^\ast\log(n)}/{\bar{m}}\right],
\end{equation*}
where $0<\rho<{1}/{2}$ and $\eta=\kappa_0^2(16-\rho)e^{M_1}/{4}$. 
\end{thm}

This theorem describes the estimation error at each iteration and provides theoretical guidance on step sizes $\eta_{\a}$, $\eta_{\W}$ and $\eta_\U$ in Algorithm \ref{alg1}. The error bound consists of two terms. The first term $\rho^t e_0$ is the computational error, which decays geometrically with the iteration number $t$ since the contraction parameter $\rho$ satisfies $0<\rho<{1}/{2}$. The second term {$\kappa_0^4e^{(1+C)M_1}\left[d\max\left\{{n}/{\bar{m}},\log(n)\right\}+{s^\ast\log(n)}/{\bar{m}}\right]$} represents the statistical error, which is related to noise in the data and does not vary with $t$. These two terms reveal an interesting interplay between the computational efficiency and statistical rate of convergence. Specifically, when the number of iterations is sufficiently large, the computational error is to be dominated by the statistical error and the resulting estimator falls within the statistical precision of the true parameters.  
In the statistical error, the term ${s^\ast\log(n)}/{\bar{m}}$ is related to estimating the sparse matrix $\W^\ast$ and the term $d\max\left\{{n}/{\bar{m}},\log(n)\right\}$ is related to estimating the low-rank matrix $\U^\ast$. The statistical error decreases with the average number of edges $\bar m$ and increases with the sparsity parameter $M_1$. 

Compared with other work on network latent space models \citep{ma2020universal,zhang2020flexible}, our theoretical analysis faces a few unique challenges. 
First, the node-topic preferential effects in $W_{ik}W_{jk}u_iu_j$ lead to an involved interplay between $\W$ and $\U$, as these effects vary across different topics and node pairs. This requires carefully bounding the error of $\W^{(t)}$ and $\U^{(t)}$ (up to rotation) separately in each step of the iteration to achieve contraction while ensuring the identifiability conditions are met. 
Second, the edge number $m_{ij}$'s vary across node pairs.
To tackle varying edge numbers, we derive a tight bound on the spectrum of random matrices with bounded moments following the techniques in \cite{bandeira2016sharp}; see Lemma S5. 
The proof of this result involves intricate technical details, and it uses large deviation estimates and geometric functional analysis techniques. The resulting bound is sharper than the matrix Bernstein inequality \citep{tropp2012user}. Using Lemma S5, we are able to improve the statistical error for low rank matrix $\U^\ast$ from $\kappa_0^4e^{(1+C)M_1}dn\log(n)$, which can be derived using the matrix Bernstein inequality under $m_{ij}=1$, to $\kappa_0^4e^{(1+C)M_1}d\max\left\{{n}/{\bar{m}},\log(n)\right\}$, which in turn relaxes the minimal signal strength condition in Assumption \ref{ass2}. Lemma S5 extends the result in \citet{lei2015consistency}, which was derived for the case of $m_{ij}=1$ using a different technique and highlights the benefit of having a greater average number of edges $\bar m$. 
Finally, the theoretical analysis is nontrivial, as it involves alternating gradient descent, orthogonal transformation, identifiability constraints, sparsity, and a non-quadratic loss function.


Based on Theorem \ref{thm1}, we can further derive the following error bounds for the estimated model parameters. 
{\begin{cor}
\label{cor1}
Under the same conditions in Theorem~\ref{thm1}, for any\\ 
$t\geq\log\left[\{{d\max\{{n}/{\bar{m}},\log(n)\}+{s^\ast\log(n)}/{\bar{m}}}\}\kappa_0^8e^{(3+C)M_1}/({C_1K{\sigma_1^\ast}^4w_{\max}^4})\right]/\log(\rho),$ 
it holds that
\begin{equation*}
\begin{aligned}
&\|\a^{(t)}-\a^\ast\|_2^2\precsim  \frac{\kappa_0^4e^{(1+C)M_1}}{K} \left[d\max\left\{\frac{1}{\bar{m}},\frac{\log(n)}{n}\right\}+\frac{s^\ast\log(n)}{n\bar{m}}\right],\\
&\|\W^{(t)}-\W^\ast\|_F^2\precsim \frac{\kappa_0^4e^{(1+C)M_1}}{w_{\max}^2} \left[d\max\left\{\frac{1}{\bar{m}},\frac{\log(n)}{n}\right\}+\frac{s^\ast\log(n)}{n\bar{m}}\right],\\
&\text{dist}^2(\U^{(t)}, \U^\ast)\precsim \frac{\kappa_0^4e^{(1+C)M_1}}{Kw_{\max}^4} \left[d\max\left\{\frac{1}{\bar{m}},\frac{\log(n)}{n}\right\}+\frac{s^\ast\log(n)}{n\bar{m}}\right],
\end{aligned}
\end{equation*}
with probability at least $1-Kn^{-1}$. 
\end{cor}}
In Corollary \ref{cor1}, the error bounds for $\a^{(t)}$ and $\U^{(t)}$ decrease with $K$, indicating that their estimation improves as the number of layers $K$ increases. 
The error bound for $W$ does not improve with $K$, as $W$ is not a common parameter shared across layers. 
All three estimation errors decrease with the average number of edges $\bar{m}$, suggesting that observing more edges between nodes leads to better estimation. Finally, the estimation error for $\U^\ast$ matches with that in standard latent space models \citep{ma2020universal} when $K=1$, $\bar m=1$ and $W_{ij}=1$ for all $i,j$.

\section{Simulation}\label{sec:sim}
In this section, we evaluate the finite sample performance of our proposed method. We also compare with some alternative solutions, and the results are collected in the supplement. Specifically, we investigate how estimation and variable selection accuracy in simulations vary with network size $n$, the number of layers $K$, edge density and the number of edges $m_{ij}$ between nodes. 
We simulate data from model \eqref{eqn:model} with parameters $\a^\ast$, $\W^\ast$ and $\U^\ast$. For $\a^\ast$, we generate its entries independently from Uniform$(a_l,a_u)$, where $a_l$ and $a_u$ together modulate the density of the network; for $\U^\ast$, we generate its rows $\u^\ast_i$'s independently from $N_d(0,\I)$, which are then scaled to ensure $\|\u^\ast_i\|_2=1$ for all $i$; for $\W^{\ast}$, we randomly select $q_0$ proportion of its entries to be nonzero and set the rest to zero; values for the nonzero entries are generated independently from Uniform(0.5, 3.5). 
We set $d=2$, $m_{ij}=m$, $q_0=0.7$ and consider $n=100, 200$, $K=10, 20, 40, 80$ and $m=1,2,4,8$. Also considered are $(a_l, a_u)=(-3.5,-1.8), (-3,-1), (-2,-1), (-1.4,-0.9)$, corresponding to an edge density of approximately 0.04, 0.08, 0.12 and 0.16, respectively.

To evaluate the estimation accuracy, we report relative estimation errors calculated as:
$$
\frac{\|\hat{\a}-\a^\ast\|_2^2}{\|\a^\ast\|^2_2}, \,\,\frac{\|\hat{ W}- W^\ast\|_F^2}{\| W^\ast\|_F^2},\,\,\min _{\R:\, \R^{\top} \R=\R \R^{\top}=\I_k}\frac{\left\|\hat{\U}-\U^\ast \R\right\|_F^2}{\left\|\U^\ast\right\|_F^2},
$$
where $\hat{\a}$, $\hat{\W}$ and $\hat{\U}$ denote the estimators from Algorithm \ref{alg1}. 
Also reported is the relative estimation error of edge probabilities $\Lambda$, calculated as 
$$
\frac{1}{K}\sum_{k=1}^{K}\frac{\left\|\psi(\hat{\Lambda}^{(k)})-\psi({\Lambda}^{ (k)^\ast})\right\|_F^2 }{\left\|\psi({{\Lambda^{(k)}}^\ast})\right\|_F^2},
$$
where ${\Lambda^{(k)}}^\ast$ is true edge probability calculated using $\a^\ast$, $\W^\ast$ and $\U^\ast$.
\begin{figure}[t!]
		\includegraphics[width=0.35\linewidth]{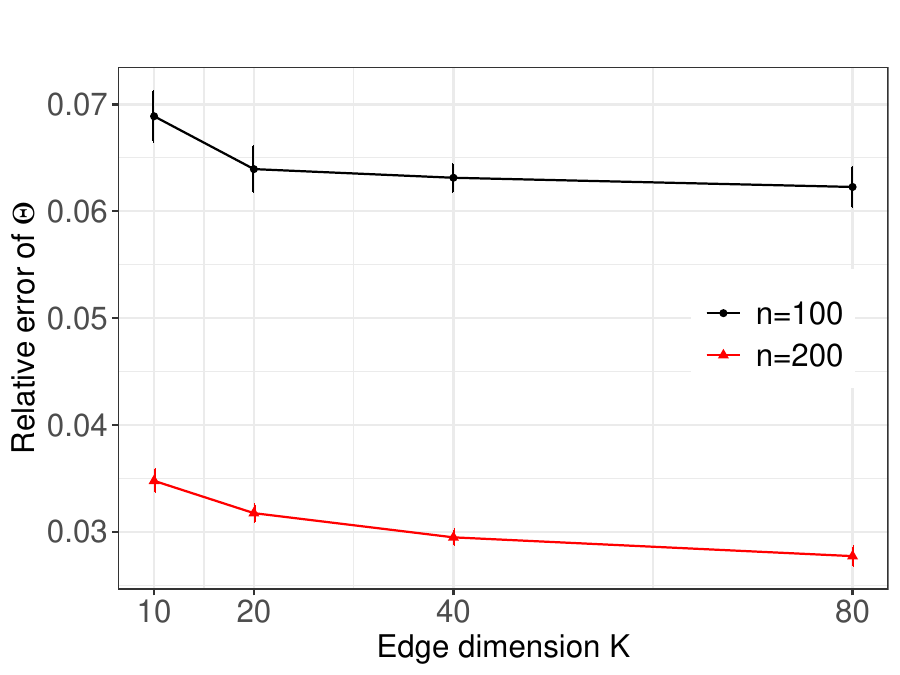}
		\includegraphics[width=0.35\linewidth]{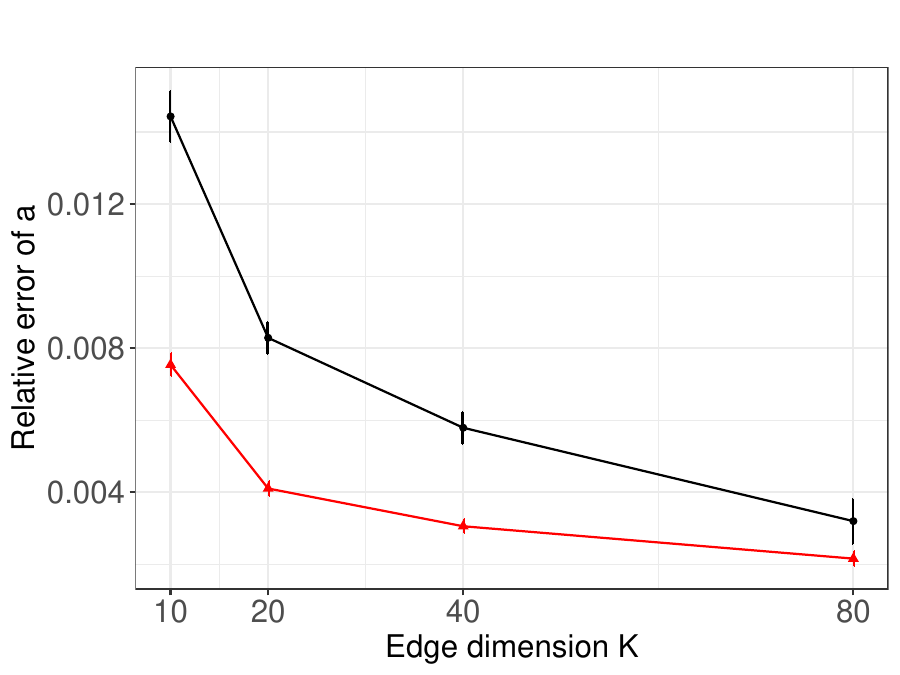}
		\centering
		\includegraphics[width=0.35\linewidth]{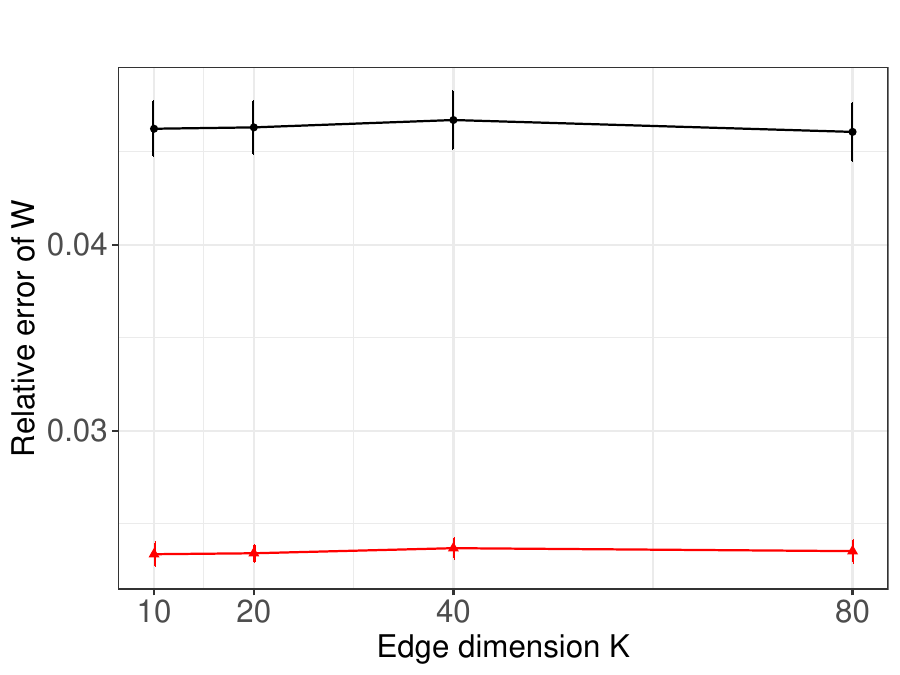}
  \centering
		\includegraphics[width=0.35\linewidth]{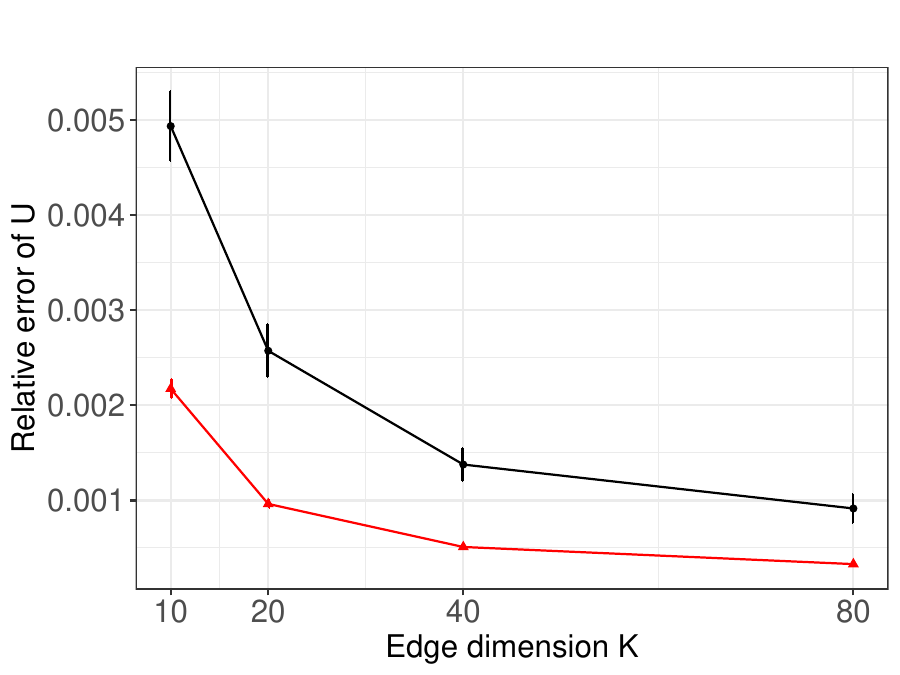}
\caption{Mean relative errors and their corresponding $95\%$ intervals under varying $n$ and $K$, while $m=1$ and edge density at 0.08. The black and red lines mark $n=100$ and $n=200$, respectively. 
 } \label{errorn100}
\end{figure}
Figures \ref{errorn100}-\ref{errordensity} report the estimation errors of ${\a}^\ast$, ${\W}^\ast$, $\U^\ast$ and $\psi({\Lambda}^{(k)}{}^\ast)$ under various settings, with 95\% confidence intervals, over 100 data replications. 
We apply the cross-validation procedure described in Section \ref{sec:est} to select the latent dimension $d$, and it consistently identifies the correct value of $d=2$.



\begin{figure}[t!]
		\centering
		\includegraphics[width=0.23\linewidth]{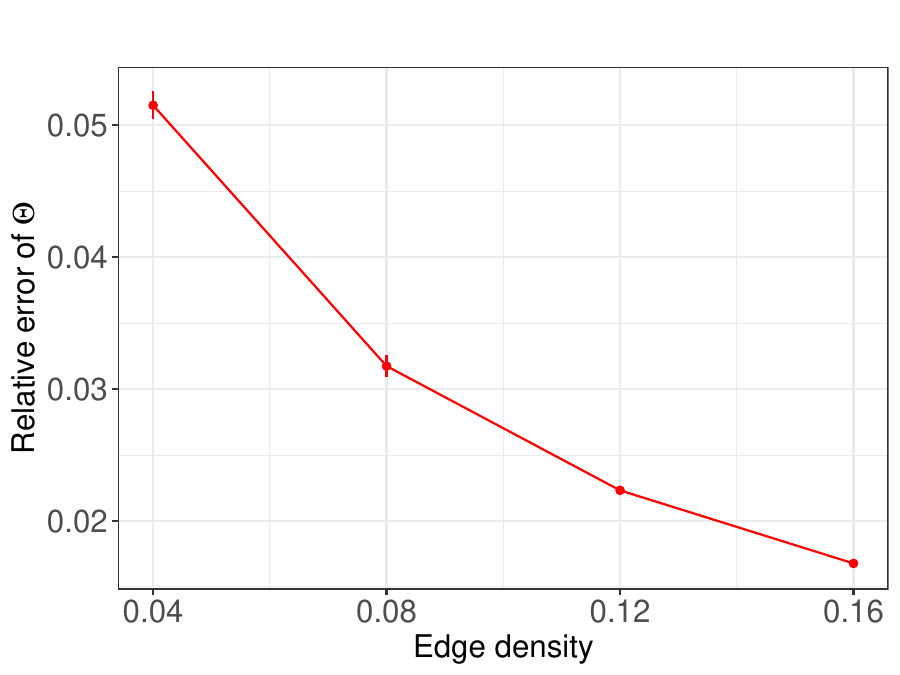}
		\centering
		\includegraphics[width=0.23\linewidth]{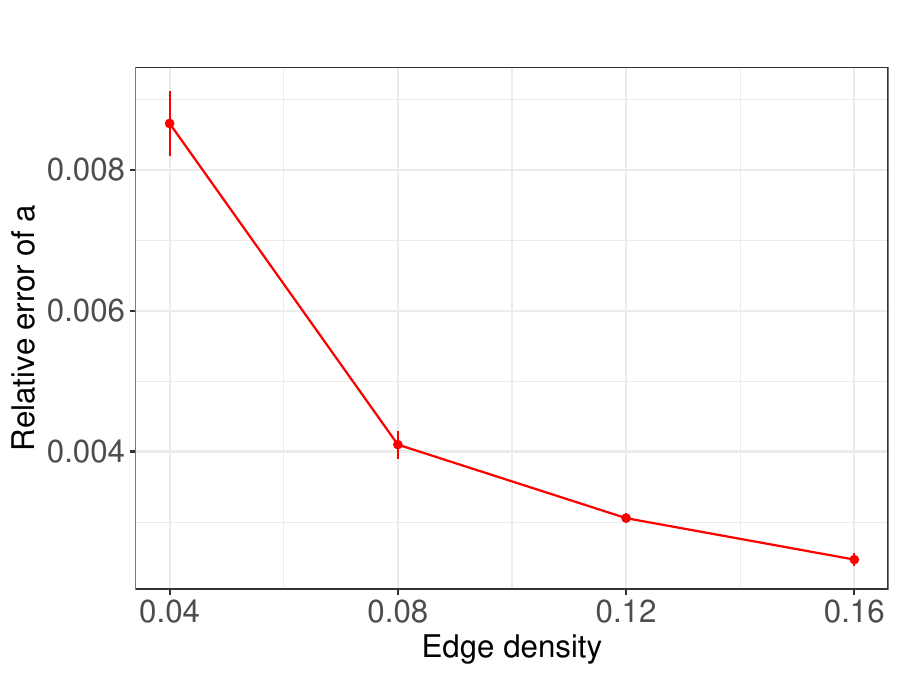}
  \centering
		\includegraphics[width=0.23\linewidth]{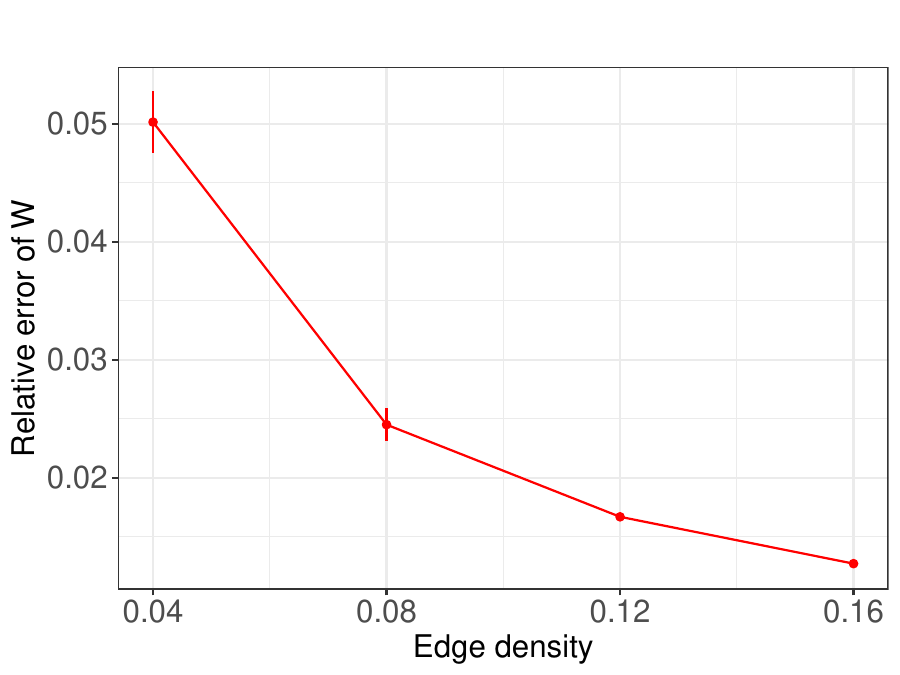}
  \centering
		\includegraphics[width=0.23\linewidth]{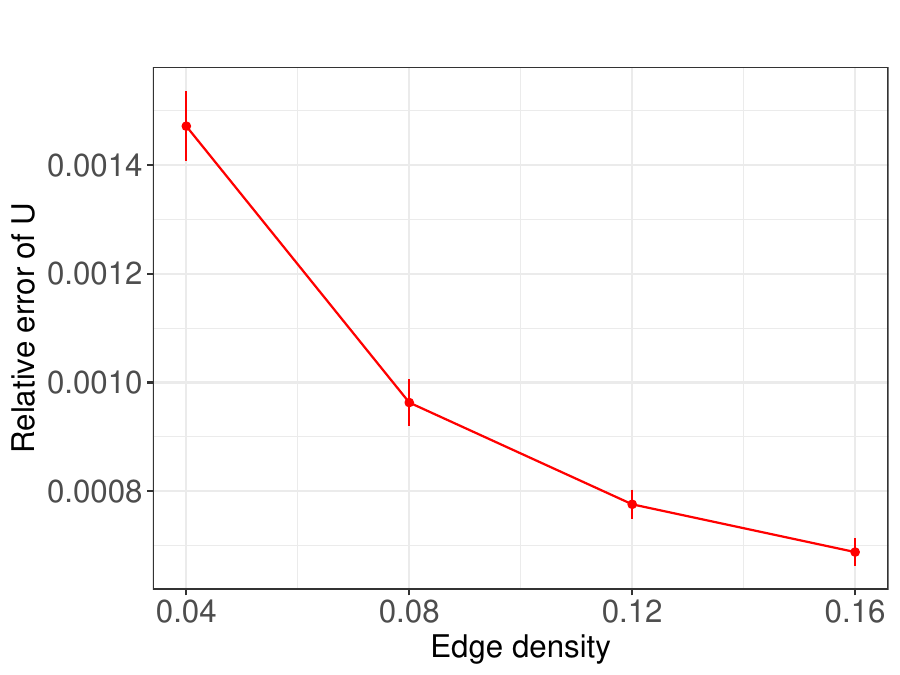}
  		\centering
		\includegraphics[width=0.23\linewidth]{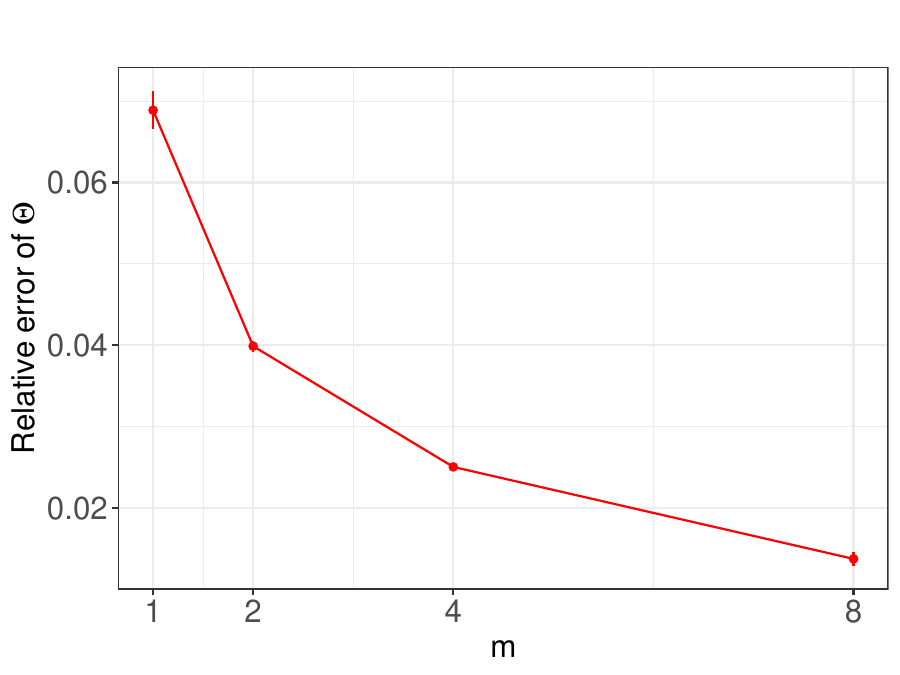}
		\centering
		\includegraphics[width=0.23\linewidth]{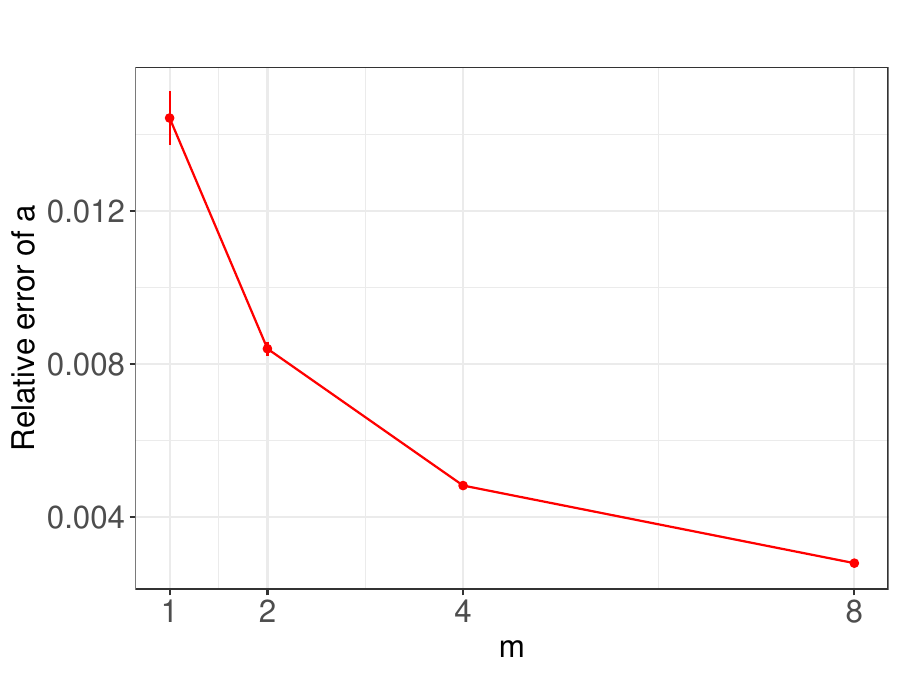}
  \centering
		\includegraphics[width=0.23\linewidth]{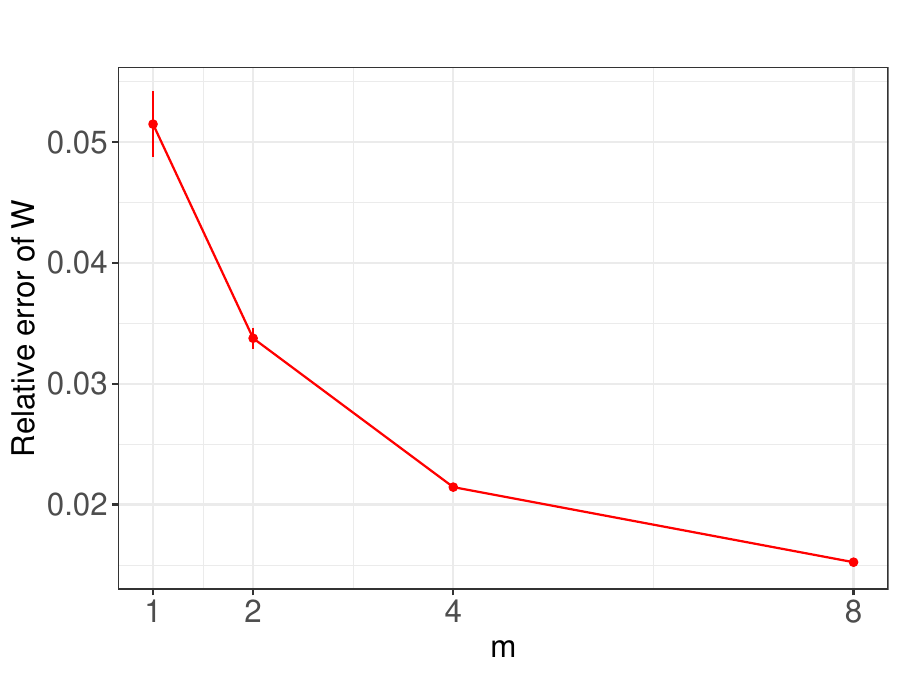}
  \centering
		\includegraphics[width=0.23\linewidth]{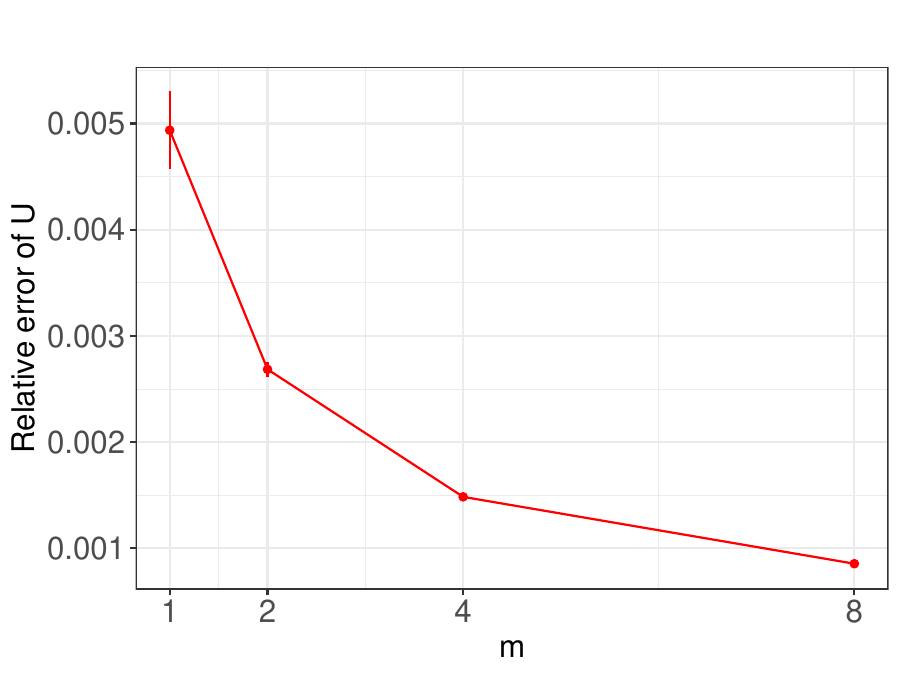}
	\caption{
Mean relative errors and their corresponding $95\%$ intervals under varying edge density while $n=200$, $K=20$, and $m=1$ (top panel), and under varying $m$ while $n=200$, $K=20$ and edge density at 0.08 (bottom panel).
 } \label{errordensity}
\end{figure}


It is seen from Figure \ref{errorn100} that estimation errors of $\a^\ast$ and $\U^\ast$ 
decrease with the network size $n$ and number of layers $K$, confirming the theoretical results in Theorem \ref{cor1}. The relative estimation error of $\W^\ast$ does not vary with $K$, as we rescale $\|\hat{ W}- W^\ast\|_F^2$ by $\| W^\ast\|_F^2$ in calculating the relative error and $\| W^\ast\|_F^2$ scales with $K$. Additionally, Figure \ref{errordensity} show that as edge density, modulated by $\a^\ast$, and the number of edges $m$
increase, the estimation errors of $\a^\ast$, $\W^\ast$ and $\U^\ast$ decrease. 

In Tables \ref{tab:tpr} and \ref{tab:tpr1}, we report the true positive rate (TPR) and false Positive rate (FPR) in estimating the nonzero entries in $\W^\ast$. The results show that the variable selection accuracy improves with $n$, $m$ and edge density. The selection accuracy remains relatively stable across different numbers of layers $K$'s, which is expected since the dimension of $\W$ increases with $K$.

\begin{table}[ht!]  
	\centering  
	\caption{True positive rate (TPR) and false positive rate (FPR) in estimating $\W^\ast$ under varying $n$, $K$, while $m=1$ and edge density at 0.08.}  
	\begin{tabular}{|c|cccc|cccc|}
	\cline{1-9}
		&\multicolumn{4}{c|}{$n=100$}&\multicolumn{4}{c|}{$n=200$}\\
	\cline{2-9}
		&$K=10$ & $K=20$& $K=40$ & $K=80$  &$K=10$ & $K=20$& $K=40$ & $K=80$ \\
		\hline
		 \multirow{2}{*}{TPR}&0.880&0.881&0.883&0.882& 0.910&0.925&0.919&0.921\\
		&(0.045)&(0.034)&(0.028)&(0.023)&(0.039)&(0.031)&(0.024)&(0.027)\\
		\cline{1-9}
		\multirow{2}{*}{FPR}&0.074&0.075&0.076&0.075&0.049&0.068&0.074&0.074\\
		&(0.039)&(0.027)&(0.022)&(0.024)&(0.022)&(0.027)&(0.028)&(0.025)\\
		\hline
  \end{tabular}
	\label{tab:tpr}
\end{table}

\begin{table}[ht!]  
	\centering  
	\caption{
 True positive rate (TPR) and false positive rate (FPR) in estimating $\W^\ast$ under varying edge density (while fixing $m=1$) and varying $m$ (while fixing edge density at 0.08), while $n=200$ and $K=20$.}
	\begin{tabular}{|c|cccc|cccc|}
	\cline{1-9}
		&\multicolumn{4}{c|}{edge density}&\multicolumn{4}{c|}{$m$}\\
	\cline{2-9}
		&$0.04$ & $0.08$& $0.12$ & $0.16$  & $1$ & $2$ & $4$ & $8$ \\
		\hline
		 \multirow{2}{*}{TPR}&0.850&0.925&0.970&0.986&0.880&0.915&0.948& 0.962\\
		&(0.045)&(0.031)&(0.007)&(0.003)&(0.045)&(0.035)&(0.021)&(0.019)\\
		\cline{1-9}
		\multirow{2}{*}{FPR}&0.128&0.068&0.032&0.015&0.074&0.066&0.058&0.035\\
		&(0.049)&(0.027)&(0.026)&(0.015)&(0.039)&(0.033)& (0.032)&(0.029)\\
		\hline
  \end{tabular}
	\label{tab:tpr1}
\end{table}



\section{Analysis of the Enron Email Network}\label{sec:real_data}

\subsection{Data description}
The Enron email corpus 
\citep{klimt}, one of the most extensive publicly available datasets of its kind, contains over 500,000 emails from 158 employees from November 13, 1998 to June 21, 2002. This dataset is released by the Federal Energy Regulatory Commission following its investigation of Enron. By analyzing this dataset with our proposed method, we offer an enriched view of the communications during one of the largest bankruptcy reorganizations in the U.S. history.

The study period can be divided into three stages, as marked by two major events. 
In February 2001, Enron's stock reached its peak and then began a dramatic decline following major sell-offs from top executives. It was later found that starting February 2001, concerns about Enron’s accounting practices were increasingly discussed internally. 
In October 2001, the company's financial scandal was publicly exposed and the Securities and Exchange Commission began an investigation into Enron’s accounting practices. 
Accordingly, we consider three stages in our analysis: the pre-decline period from November, 1998 to February, 2001; the decline and pre-bankruptcy period from February, 2001 to October, 2001; and the bankruptcy and post-bankruptcy period from October, 2001 to June, 2002. 


\begin{table}[t!]
\renewcommand{\arraystretch}{0.8} 
\footnotesize
\centering
\caption{Extracted topics from the Enron email corpus. Within each category, topics are sorted in descending order based on their frequency of occurrence across all emails.}
\label{tab:LDA}
\begin{tabular}{@{}>{\raggedright\arraybackslash}p{7cm} >{\raggedright\arraybackslash}p{7cm}@{}}
\toprule
\textbf{Category} & \textbf{Topics} \\
\midrule
\multirow{8}{7cm}{\textit{Legal and Regulatory Affairs}} 
& legal and contractual issues \\
& legal department contacts \\
& compliance and pipeline management \\
& LNG financing opportunities \\
& migration issues \\
& BHP market inquiry \\
& customer service inquiries \\
& article reviews \\
\midrule
\multirow{8}{7cm}{\textit{Energy Markets and Operations}} 
& California energy crisis \\
& NYMEX website issues \\
& energy market strategies \\
& energy index management \\
& accessing westpower desk \\
& energy portfolio management \\
& draft review process \\
& document management issues \\
\midrule
\multirow{8}{7cm}{\textit{Administrative Coordination}} 
& scheduling meetings \\
& task or role reassignment \\
& communication coordination \\
& document review process \\
& time zone conversion \\
& recognition and support \\
& variance methodology discussion \\
& project collaboration tools \\
\midrule
\multirow{6}{7cm}{\textit{Corporate Strategy and Projects}} 
& strategy session updates \\
& financial data publication \\
& Vince Kaminski project \\
& document management issues \\
& election concession parodies \\
& fair trade opinions \\
\midrule
\multirow{3}{7cm}{\textit{Technology and Tools}} 
& blackberry handheld devices \\
& password security procedures \\
& communication issues \\
\midrule
\multirow{11}{7cm}{\textit{Social and Interpersonal Communication}} 
& personal relationships and communication \\
& informal workplace conversations \\
& miscommunication apologies \\
& taste and acquired preferences \\
& game interactions \\
& Keneally's social night \\
& sailing lessons in Australia \\
& wine retail pricing \\
& socializing and drinks \\
& congratulations and well wishes \\
& family communications network \\
\midrule
\multirow{2}{7cm}{\textit{Miscellaneous and Culture}} 
& horoscope and relationships \\
& independence day plans \\
\bottomrule
\end{tabular}
\end{table}

First, emails are preprocessed by removing punctuation, lemmatization, stopwords, and documents with less than 5 words. Then, we consider a transformers-based topic modeling method \citep{grootendorst2022bertopic} to extract latent topics from the Enron email dataset. In the procedure, texts are first embedded using pre-trained transformer `ll-MiniLM-L6-v2', and then go through a dimension reduction step via uniform manifold approximation and projection (UMAP) \citep{mcinnes2018umap}. The embeddings are clustered to identify topics, where the theme of each topic is extracted using cluster-based TF-IDF \citep{sparck1972statistical}, and then fine-tuned using the GPT-4o Mini language model. This results in 47 well-defined and distinct topics (see Table \ref{tab:LDA}}). More details of this data processing procedure can be found in Section S8. 

Our analysis focuses on the emails of 154 employees whose roles and departments are documented in the dataset. 
We begin by applying topic modeling to the full set of emails from all three stages to extract topics. Each email is then assigned one or more topics based on its content. To construct the network for each stage, we proceed as follows: for each pair of employees, we aggregate all emails between them in a given stage and construct a length-$K$ binary vector indicating the presence or absence of each topic in their communication. This process results in an undirected, multi-layer network for each stage with $n = 154$ nodes (employees) and $K = 47$ layers (topics). 
The presence of an edge in layer $k$ is denoted by $y_{ij}^{(k)} = y_{ji}^{(k)} = 1$ if any email between users $i$ and $j$ is assigned topic $k$, and $y_{ij}^{(k)} = y_{ji}^{(k)} = 0$ otherwise. 
This aggregation approach helps to reduce the sparsity of the network and facilitates comparison with other multi-layer network analysis methods. The edge densities of the resulting networks for stages 1, 2, and 3 are 0.23\%, 0.22\%, and 0.13\%, respectively. 



\subsection{Alternative approaches and link prediction}
We consider three alternative approaches when analyzing the Enron data:
\begin{itemize}
\item \texttt{Separate}: This method fits a separate latent space model to each layer, that is, 
$\Lambda^{(k)}=\a^{(k)}  1_n^T +  1_n {\a^{(k)}}^\top+\U^{(k)}{\U^{(k)}}^{\top}$ for $k \in [K]$.
\item \texttt{\texttt{Multiness}} \citep{macdonald2022latent}: This method includes a common latent structure across layers and a separate latent structure for each individual layer, written as $\Lambda^{(k)}= V \I_{p_0, q_0}  V^{\top}+\U^{(k)} \I_{p_{k}, q_{k}} {\U^{(k)}}^{\top}$ for $k \in [K]$, where $ V \in \mathbb{R}^{n \times d_0}$ is the matrix of common latent positions, $\U^{(k)} \in \mathbb{R}^{n \times d_{k}}$ collects the individual latent positions for layer $k$, and 
$
\I_{p, q}=\left(\begin{array}{cc}
\I_p & 0 \\
0 & -\I_q
\end{array}\right)$. 

\item \texttt{FlexMn} \citep{zhang2020flexible}: This method considers 
layer-specific degree heterogeneity $\a_k$, and a ommon latent position ${U}$ across layers with a layer-specific scaling matrix ${\Lambda^{(k)}}$, written as $
\Lambda^{(k)}=\a^{(k)} 1_n^{\top}+1_n {\a^{(k)}}^{\top}+\U {\Lambda}^{(k)}\U^{\top}$, for $k \in [K]$, where $\a^{(k)} \in \mathbb{R}^n$, $\U \in \mathbb{R}^{n \times d}$, and ${\Lambda}^{(k)} \in \mathbb{R}^{d \times d}$.
\end{itemize}

\begin{figure}[t!]
  		\subfigure[stage 1]{
		\centering
		\includegraphics[width=0.3\linewidth]{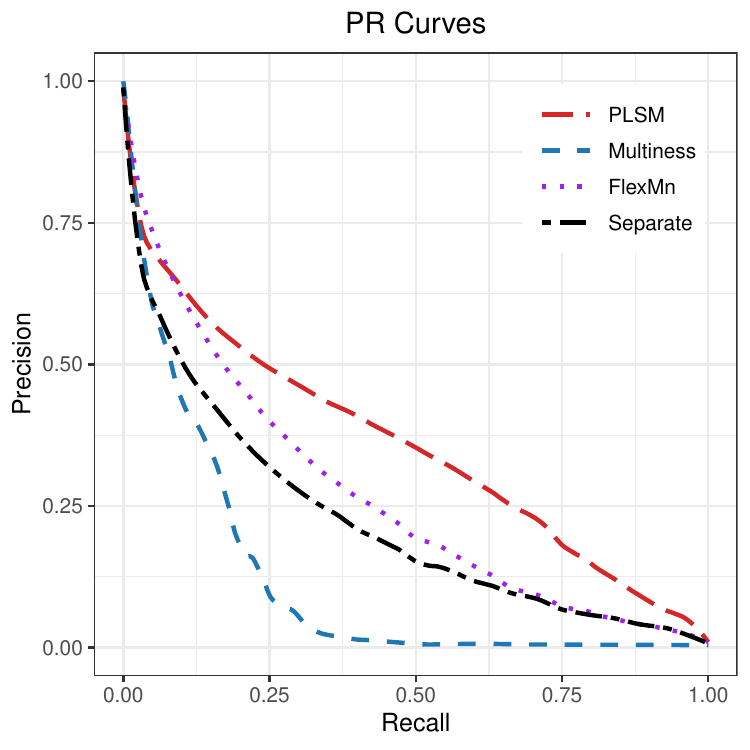}}
  \subfigure[stage 2]{
		\centering
		\includegraphics[width=0.3\linewidth]{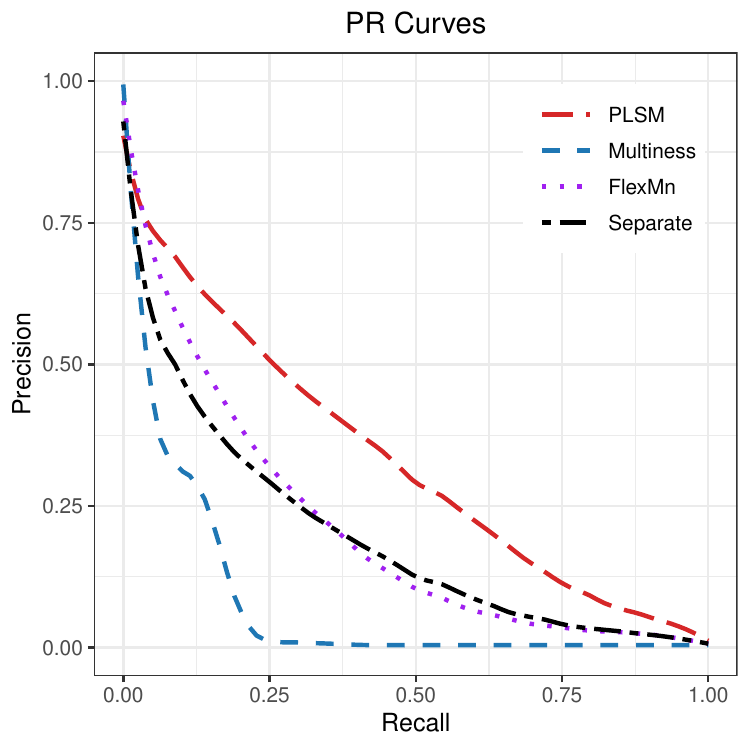}}
   \subfigure[stage 3]{
		\centering
		\includegraphics[width=0.3\linewidth]{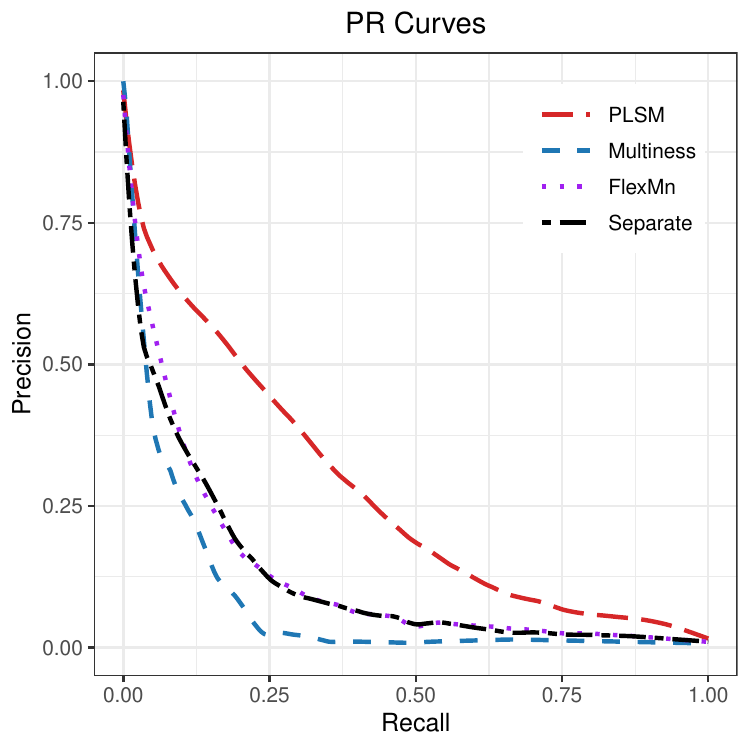}}
	\caption{Precision-Recall curves from different methods for out-of-sample link prediction.} \label{pr}
\end{figure}

To compare the performance of above methods in link prediction, we randomly remove 20\% entries from each layer and treat them as missing data. We then apply \texttt{PLSM}, \texttt{Multiness}, \texttt{FlexMn}, and \texttt{Separate} to the remaining entries and use the fitted model to predict link probabilities for the missing entries. This procedure is repeated 100 times. To ensure a fair comparison, edge cross-validation is used in selecting the latent space dimension for all methods.
Figure \ref{pr} shows the average precision-recall curves from all four methods. It is seen that \texttt{Separate} does not perform well as it cannot borrow information across different layers; \texttt{Multiness} might have suffered from over-fitting as there is a large number of sparse layers in each of the three networks. \texttt{FlexMn} assumes a shared latent position across layers with layer-specific scaling, which may limit its flexibility in capturing heterogeneity in node-topic preferences. \texttt{PLSM} enjoys the best performance among all methods.
Comparisons of these methods in simulations are included in Section S7.


\begin{table}[t!]
\renewcommand{\arraystretch}{1.15}
\centering
\small
\caption{Descriptions of main departments in the Enron email dataset.\label{tab:depart}}
\begin{tabular}{>{\raggedright\arraybackslash}p{2cm} >{\raggedright\arraybackslash}p{6cm} >{\raggedright\arraybackslash}p{6.5cm}}
\toprule
\textbf{Acronym} & \textbf{Full Name} & \textbf{Function} \\\midrule
Gas & Gas Divisions & Natural gas trading \\
Legal & Legal Division & Legal and compliance \\
ETS & Enron Transportation Services & Logistics and infrastructure related to energy transportation \\
RGA & Regulatory and Government Affairs & Communication with government and regulatory agencies \\
EWS & Enron Wholesale Services & Wholesale trading operations and financial products \\
\bottomrule
\end{tabular}
\end{table}

\subsection{Estimation results from \texttt{PLSM}}
\begin{figure}[t!]
    \centering
\includegraphics[width=0.95\linewidth]{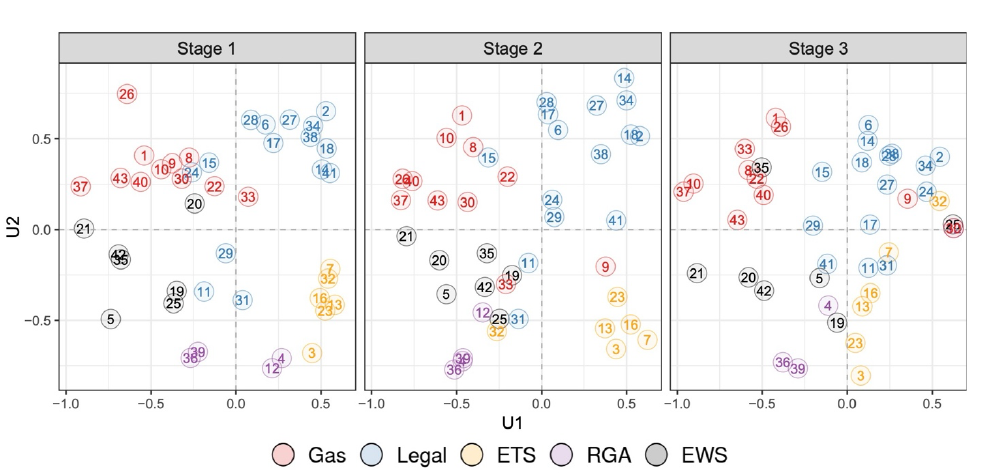}
    \caption{Estimated latent positions $U$ by \texttt{PLSM} for stages 1–3 (left to right).
    }\label{commonu}
\end{figure}

We apply our proposed method, \texttt{PLSM}, to the three networks from stages 1-3 and use edge cross-validation to select the latent dimension $d$ and the sparsity $s$. 
Edge cross validation selects a latent dimension of 6 for stage 1, 5 for stage 2, and 7 for stage 3. 
The proportion of non-zero entries in $\W$ is selected as 0.85, 0.55 and 0.6  for stages 1, 2, and 3, respectively. To facilitate visualization, we focus on 43 employees from five major departments who hold positions at or above the director level. 
The description of the departments can be found in Table \ref{tab:depart}. 
The estimated latent position $\u_i$'s from \texttt{PLSM} are of unit length, placing them on a $K$-dimensional sphere. To ensure the comparability of latent positions $U$ across three stages, we employ Procrustes analysis \citep{gower1975gpa} to align them in a common coordinate system and visualize the first two dimensions; see Figure \ref{commonu}.
Nodes with closer latent positions are more likely to engage in communication.
The clustering pattern of nodes in stages 1 and 2 shows that before bankruptcy, executives in different departments function relatively autonomously. In stage 3, there is an increase in cross-departmental communications. In stage 3, we also observe an increase in communications involving the legal department, as nodes from the legal department move closer to others. 

We also observe some interesting individual-level patterns. 
For instance, Chris Germany (node \#9), Manager of Gas Trading, was central in communications within GAS in stage 1. In stages 2-3, his interactions had a noticeable shift towards ETS and legal departments, reflecting a potential change in responsibilities. 
Susan Scott (node \#32), Counsel for the ETS department, moved closer to the RGA department in stage 2, and shifted towards the legal department in stage 3, suggesting a growing involvement in regulatory and legal discussions.


\begin{figure}[t!]
    \centering
\includegraphics[width=0.9\linewidth]{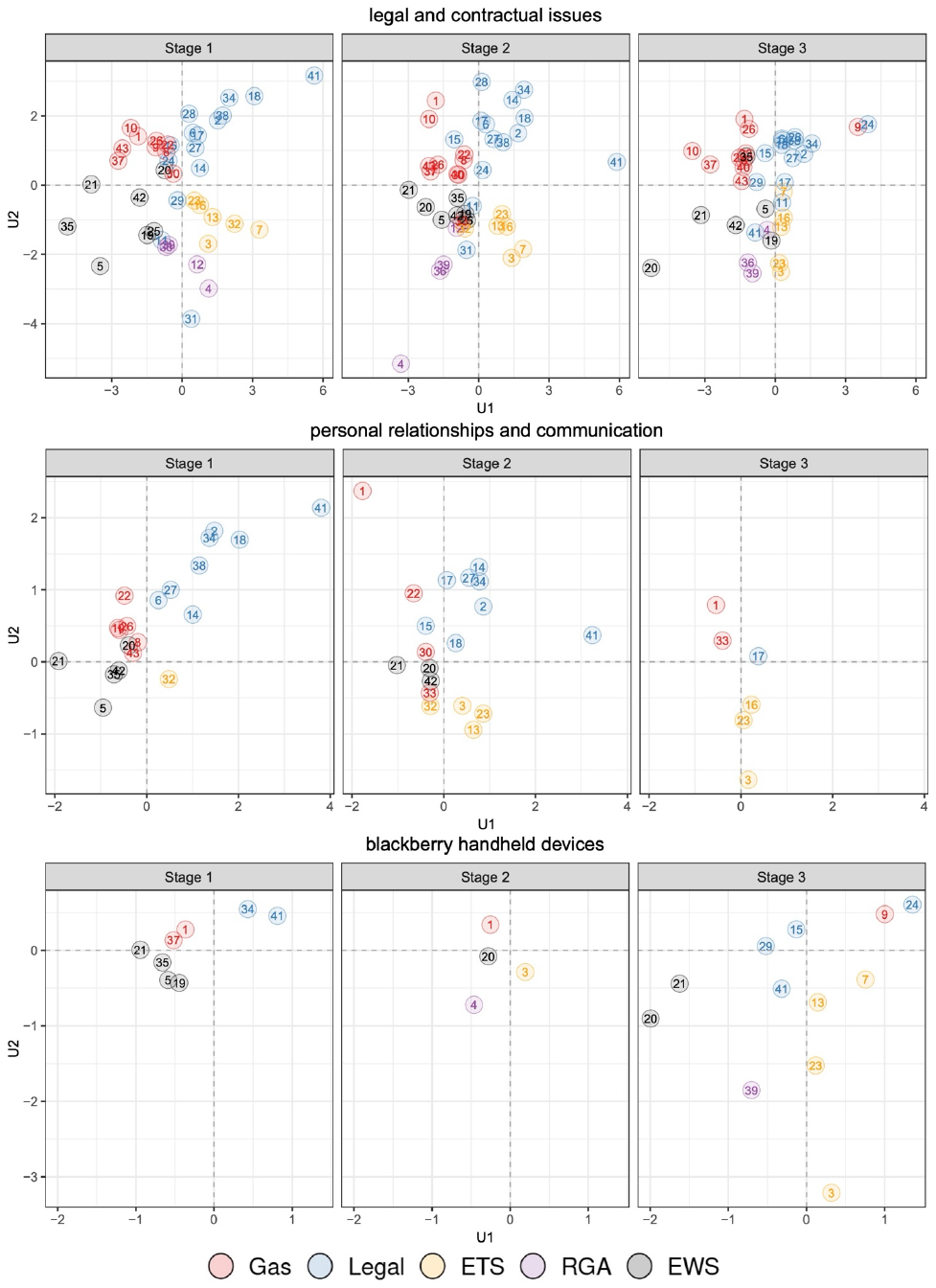}
    \caption{Estimated preferential latent positions $W_{ik}\u_i$'s by \texttt{PLSM} on selected topics for stages 1–3 (left to right). For clearer visualization, individuals who are close to (0,0) are removed (squared $\ell_2$ distance is less than 0.1).
    }\label{Wu}
\end{figure}

In addition to the latent positions $\u_i$'s, our proposed method also provides preferential latent positions calculated as $W_{ik} {u}_i$'s. In particular, the direction of $W_{ik} {u}_i$ is the same as $\u_i$, while its length $\|W_{ik} \u_i\|_2 = W_{ik}$ characterizes the activeness of node $i$ on topic $k$. Figure \ref{Wu} shows the first two dimensions of the matrix $[W_{1k} \u_1, \ldots, W_{nk} \u_n]^\top$ for three selected topics, ``legal and contractual issues", ``personal relationships and communication" and ``blackberry handheld devices".
It is seen that, on the topic of ``legal and contractual issues", members of the legal department moved closer to other departments from stage 1 to stage 3. For instance, Mark E. Taylor (node \#41), Vice President and General Counsel of the legal department, and Jeff Dasovich (node \#4), a Director in the RGA department, had increasing communication in stage 3, although they had distinct positions in stages 1-2.
Participation in the ``personal relationships and communication" topic dropped sharply across all departments, especially in EWS and Legal, as attention shifted towards crisis response. This decline in personal communication reflects the disruption of Enron’s workplace culture amid the escalating crisis. 
One exception is John Arnold (node \#1), the VP of gas department, 
who had activity on this topic in stages 2-3. 
On the ``BlackBerry handheld devices'' topic, we observe a sudden increase in activity in stage 3 among Shelley Corman (node \#3), VP of ETS, James D. Steffes (node \#39), VP of Government Affairs, and Michelle Lokay (\#23), Director of ETS, showing a surge in mobile-based communication among senior leadership. This suggests that, during the crisis, mobile devices became a important channel for rapid coordination and decision-making.

\begin{figure}[t!]
\centering
{\includegraphics[width=0.9\linewidth]{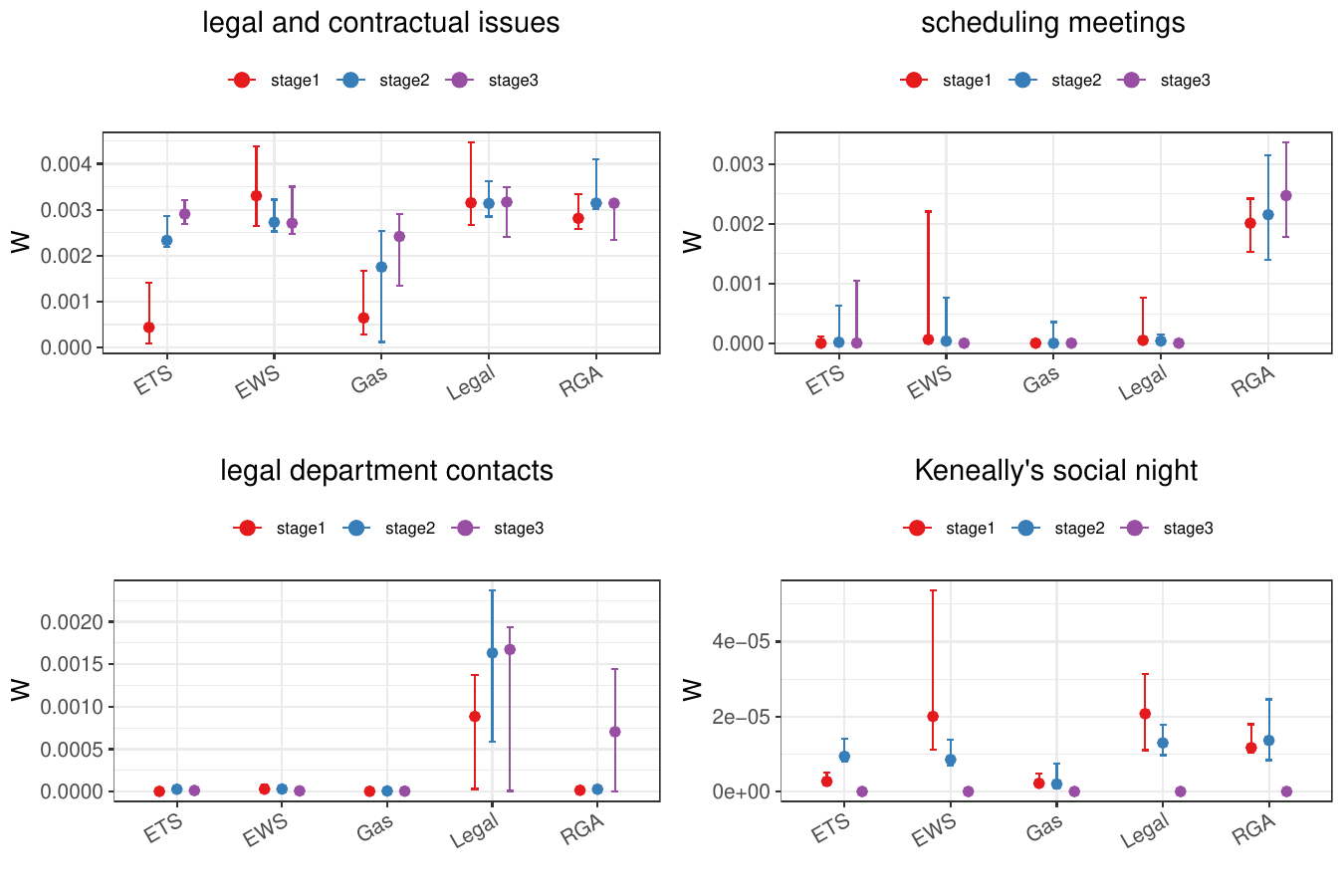}}
\caption{Average topic weights ($W_{ik}$) for 4 selected topics by department and stage, with error bars representing the first and third quartiles.} \label{W}
\end{figure}

Figure \ref{W} presents the scaled weights of four selected topics across five departments over three stages. For each department, the bars represent the interquartile range of the weights within this department, and the circle marks the average weight within this department. The patterns highlight how each department shifted its communication focus throughout the three stages, reflecting changes in priorities and responses to the crisis. 
Across Enron's organizational decline, we observe a notable increase in attention to the topic ``legal and contractual issues" within two core business departments, ETS and Gas, 
 showing that legal issues became more closely tied to business activities as the company moved towards bankruptcy. 
In contrast, departments with legal or regulatory functions, EWS, Legal, and RGA had consistent engagement with this topic throughout all stages. 
This pattern suggests that legal discussions were spreading into business-focused departments while core legal functions remained active across the organization.
In Stage 3, we observe an increased engagement from the RGA (Regulatory and Government Affairs) department with the topics ``Legal Department Contacts" and ``Scheduling Meetings." 
This reflects that RGA had a growing responsibility in helping with legal disclosures, supporting legal actions, and managing communication between departments and outside agencies during and after the bankruptcy. 

The topic ``Keneally’s social night" refers to casual conversations about after-work gatherings at Keneally’s, a local bar popular among Enron employees. 
These interactions show how coworkers connected outside the formal office environment. 
Mentions of ``Keneally’s social night" dropped sharply across all departments, especially in EWS, Legal, and RGA, as these groups became more involved in crisis response and legal matters. The disappearance of informal topics like this one shows how the growing crisis disrupted Enron’s workplace culture, and marked the breakdown of the company’s social environment.

\section{Discussion}\label{dis}
This work introduces a preferential latent space modeling framework for networks with rich textual information. To incorporates texts into the analysis of networks, we use transformer-based word embeddings together with a topic extraction process that produces interpretable topic-aware embedding for text associated with each edge in the network. We formulate a new and flexible preferential latent space model that can offer direct insights on how node-topic preferences modulate edge probabilities. We establish identifiability conditions for the proposed model and tackle estimation using a projected gradient descent algorithm. We further establish theoretical guarantee by providing the non-asymptotic error bound for the estimator from each step of the algorithm. 

Our newly proposed preferential latent space model can be used to model other multi-layer networks, particularly when there are node-layer heterogeneity. Examples include multi-layer social networks, where users interact on different platforms, such as Facebook, Twitter, Instagram, with varying levels of engagement, and international trade networks, where countries trade on different products with varying levels of demands.

Future work can extend our model in several directions. One natural extension is to directed networks, where the direction of communication (e.g., sender or receiver) carries important information. Another is to temporal networks, which incorporates the timing of interactions over time. Additionally, our work can also be extended to incorporate node-level and/or edge-level covariates. This direction can be developed following the approach in \citet{ma2020universal}. We leave these directions to future research.

\bibliographystyle{chicago}
\begingroup
\baselineskip=17.5pt
\bibliography{LSM}
\endgroup
\end{document}